\documentclass[longauth]{aa}  
\usepackage{amsmath,amssymb,amsfonts}
\usepackage{graphicx}
\usepackage{txfonts}
\usepackage{enumerate} 
\usepackage{colordvi} 
\usepackage{bm}
\usepackage{epsfig}
\usepackage{hyperref}
\usepackage[dvipsnames]{xcolor}
\usepackage{color, colortbl}
\usepackage[T1]{fontenc} 
\usepackage{braket}
\usepackage{euclid}
\usepackage{placeins}
\usepackage{tabularx}
\usepackage{ulem}
\usepackage{orcidlink}

\definecolor{amethyst}{rgb}{0.6, 0.4, 0.8}

\definecolor{amaranth}{rgb}{0.9, 0.17, 0.31}
\definecolor{forestgreen(web)}{rgb}{0.13, 0.55, 0.13}
\definecolor{lavender(web)}{rgb}{0.9, 0.9, 0.98}
\definecolor{cosmiclatte}{rgb}{1.0, 0.97, 0.91}
\definecolor{jonquil}{rgb}{0.98, 0.85, 0.37}
\definecolor{khaki(x11)(lightkhaki)}{rgb}{0.94, 0.9, 0.55}
\definecolor{thistle}{rgb}{0.85, 0.75, 0.85}

\newcommand{\lcdm}{\ensuremath{\Lambda\mathrm{CDM}}}

\newcommand{\de}{\mathrm{d}}

\newcommand{\orcid}[1]{}

\usepackage[nameinlink,noabbrev]{cleveref}
\Crefname{equation}{Eq.}{Eqs.}
\Crefname{section}{Sect.}{Sects.}
\Crefname{figure}{Fig.}{Figs.}
\crefname{equation}{Equation}{Equations}
\crefname{section}{Section}{Sections}
\crefname{figure}{Figure}{Figures}

\begin{document}
\title{Forecast on the generalised dark matter properties from a Euclid-like survey}

\author{Ziad Sakr\orcidlink{0000-0002-4823-3757}\thanks{\email{zsakr@irap.omp.eu}}\inst{\ref{aff1},\ref{aff2},\ref{aff3}}
\and
Jessica N. López-Sánchez\orcidlink{0000-0002-4489-7820}\thanks{\email{lopez@fzu.cz}}\inst{\ref{aff4}}}

\institute{Instituto de Física Teórica UAM-CSIC, Campus de Cantoblanco, 28049 Madrid, Spain \label{aff1}
\and
Institut de Recherche en Astrophysique et Plan\'etologie (IRAP), Universit\'e de Toulouse, CNRS, UPS, CNES, 14 Av. Edouard Belin, 31400 Toulouse, France \label{aff2}
\and
Universit\'e St Joseph; Faculty of Sciences, Beirut, BP-11514, Lebanon\label{aff3}
\and 
Institute of Physics of the Czech Academy of Sciences, Na Slovance 1999/2, 182 00 Prague, Czech Republic\label{aff4}
}

\date{\today}

\authorrunning{Sakr et al.}

\titlerunning{Euclid-like: forecasts on generalized dark matter}


  \abstract
  {
The Stage~IV \textit{Euclid} mission will deliver spectroscopic galaxy redshifts together with photometric positions and shapes, enabling cosmological analyses through spectroscopic galaxy clustering (GCsp), photometric galaxy clustering (GCph), weak-lensing cosmic shear (WL), and their cross-correlation (XC). These probes will provide strong constraints on extensions of the concordance \(\Lambda\)CDM model, including scenarios in which the properties of cold dark matter are modified. In this work we forecast the constraining power of a Euclid-like survey on the Generalised Dark Matter (GDM) parameters \(w_{\rm gdm}\) and \(c^{2}_{s,{\rm gdm}}\).
Our analysis builds upon and extends previous forecasting pipeline used for standard cold dark matter. For GCsp, we adopt a semi-analytic nonlinear RSD model, with free terms for each bin, that also includes Fingers-of-God and BAO damping. For the photometric probes, we compute the nonlinear GDM matter power spectrum using dedicated simulations, and we modify the lensing and clustering window functions as well as the intrinsic-alignment prescription according to our model. We consider several survey configurations characterized by different maximum multipole or wavenumber cuts to assess the constraining power of each setup. Within each configuration, we additionally explore three fiducial values of \(\sigma_8\) motivated by current CMB and low-redshift measurements.
In an optimistic setting, for fiducial values \(\sigma_8 \simeq 0.81\) and \(\sigma_8 \simeq 0.77\), we find relative errors of \(4.01\%\) (GCsp), \(5.01\%\) (GCph+WL+XC), and \(1.96\%\) (all probes) on \(c^{2}_{s,{\rm gdm}}\), and \(3.26\%\) (GCph+WL+XC) and \(1.85\%\) (all probes) on \(w_{\rm gdm}\). For a lower fiducial value \(\sigma_8 \simeq 0.67\), that could strongly disfavor $\Lambda$GDM,  we find constraints of \(5\%\) (GCsp), \(5\%\) (GCph+WL+XC), and \(2.45\%\) (all probes) on \(c^{2}_{s,{\rm gdm}}\), and \(3.43\%\) (GCph+WL+XC) and \(2.04\%\) (all probes) on \(w_{\rm gdm}\). We also found that, when combining all probes, whether in the pessimistic or optimistic settings, a Euclid-like survey will be able to disentangle between the three aforementioned \(\sigma_8\) scenarios. 
These results show that a Euclid-like survey will be able to constrain the GDM parameters at the percent level and distinguish between different normalisations of the matter fluctuations, thereby providing a sensitive test of extensions to cold dark matter.  
   }


   \maketitle

\section{Introduction}\label{sec:intro}

Nowadays, most cosmological data suggest a cosmic expansion history consistent with a flat geometry and the presence of dark energy—usually in the form of a cosmological constant, $\Lambda$, to explain the accelerating expansion of the Universe. It is also assumed that large-scale structure formed thanks to a cold dark matter (CDM) component, described as a nonrelativistic fluid whose influence is purely gravitational. However, there is no fundamental theoretical reason to exclude the possibility of a non-zero dark matter EoS parameter. In fact, despite the fact that the $\Lambda$CDM model has been successful in describing the large-scale structure, it still faces several related challenges, with respect to the Cold Dark Matter (CDM) component, since the observed properties of halos deviate from the predictions of the CDM framework \cite{Weinberg2015}. 

In order to alleviate some of the issues of CDM, various dark matter (DM) candidates have been proposed, such as unstable DM \cite{Doroshkevich:1989bf}, warm DM \cite{Colin2000,Bode2001}, fuzzy DM \cite{Hu2000}, condensate DM \cite{Sin1994,Goodman2000}, and decaying DM \cite{Cirelli2012} (see also \cite{Bullock2017} for a review or \cite ). Most of these models can be effectively modeled within the Generalized Dark Matter (GDM) framework, which describes the dark matter fluid with a  positive pressure parametrized by a background equation of state $w_{\rm gdm}$ and a non-vanishing sound speed $c_{\rm s,gdm}^2$, as first introduced in \cite{Hu1998}. This approach has motivated several researchers to investigate whether observational data support non-cold DM (i.e., a non-zero EoS parameter of DM) \cite{Muller2005,Piattella2012,Calabrese2011,Beutler2014,Audren2014}. For instance, to test the warmness of DM, \cite{Muller2005} investigated the DM EoS parameter using cosmic microwave background (CMB), type Ia supernova (SN) and large scale structure data with zero adiabatic sound speed and no entropy production. \cite{Kumar2019} explores the non-zero DM EoS parameter using CMB, baryon acoustic oscillations (BAO) and SN data in the DE and DM interaction framework, providing evidence at approximately $1\sigma$ level. In particular, this has boosted research into scenarios involving non-zero dark matter EoS \cite{Kumar2019,RoyChoudhury2020,Haridasu2021}. For example, \cite{Kumar2019b} investigated the potential deviations from CDM and found that the current observational data favor a non-zero dark matter EoS parameter at approximately the $2\sigma$ confidence level.

In this work, we forecast the properties of the GDM model using Stage-IV-like surveys such as \cite{DESI2016}, Rubin/LSST \cite{Marshall2017}, \cite{Dore2019}, and \cite{SKA2018}. We especially focus on a Euclid-like survey configuration \cite{Laureijs2011} that will include a photometric survey (measuring positions and shapes of galaxies for tomographic weak-lensing and angular clustering analyses) and a spectroscopic survey (providing precise radial galaxy positions for 3D clustering), allowing for a combined large scale structures (LSS) + galaxy weak lensing correlations analysis that yields stringent constraints on GDM parameters.

It is important to emphasize that the original GDM framework allows for negative values of the pressure and non-vanishing viscosity term $c_{\rm vis,gdm}^2$. The implications of this modeling for the background expansion and linear perturbations have been extensively studied \cite{Kopp2016,Kunz2016,Thomas2016,Kopp2018,Tutusaus2018,Thomas2019}. However, we do not include the viscosity term here because, in linear order, it degenerates with the sound speed $c_{\rm s,gdm}^2$. 
Although a time-dependent treatment could in principle break this degenerency, current data are insufficient to constrain any temporal evolution of the GDM parameters \cite{Kopp2018}. While future data may provide tighter bounds, the nonlinear implementation of this additional parameter is not currently available nor made so by us due to the additional complexity needed to properly include it. Therefore, in this study, we assume constant values for both the dark matter equation of state and sound speed parameters.


\section{Theoretical Background}\label{sec:fR}
In this work, we consider the GDM model, characterized by a background equation of state parameter $w_{\text{gdm}}$ and a non-vanishing sound speed $c_{\rm s,gdm}^2$. We assume that DM only interacts gravitationally with the other components and that all fluids satisfy the standard continuity equation $\nabla_{\nu}T_i^{\mu\nu}=0$, where $T_i^{\mu\nu}$ is the stress-energy tensor. For a perfect fluid, this term reads as
\begin{equation}
  T_i^{\mu\nu}=(\rho_i c^2+P_i)u^{\mu}u^{\nu}+P_ig^{\mu\nu}\,,
\end{equation}
where $g^{\mu\nu}$ the metric is the metric; $\rho_i$, $P_i$ and $u_i$ are the density, the pressure, and the four-velocity of each fluid, respectively.

Contracting the continuity equation once with $u_{\mu}$ and once with the projection operator $h_{\mu\alpha}=g_{\mu\alpha}+u_{\mu}u_{\alpha}$, one obtains the relativistic expressions for the continuity equations and the Euler equations, respectively:
\begin{eqnarray}
  \frac{\partial\rho_i}{\partial t} + \nabla_{\vec{r}}\cdot(\rho_i\vec{v}_i) + 
  \frac{P_i}{c^2}\nabla_{\vec{r}}\cdot\vec{v}_i = 0\,, \label{equ:cnpert} \\
  \frac{\partial\vec{v}_i}{\partial t} +  (\vec{v}_i\cdot\nabla_{\vec{r}})\vec{v}_i+
  \nabla_{\vec{r}}\Phi+\frac{\nabla_{\vec{r}}P_i}{\rho_i+P_i/c^2}=0\,. \label{equ:enpert}
\end{eqnarray}
Here $\vec{v}_i$ is the three-dimensional velocity of each species, $\Phi$ the Newtonian gravitational potential, and $\vec{r}$ denotes the physical coordinates.

The $00$-component of Einstein's field equations gives the relativistic Poisson equation
\begin{equation}\label{equ:pnpert}
 \nabla_{\vec{r}}^2\Phi = 4\pi G\sum_k\left(\rho_k+\frac{3P_k}{c^2}\right)\,
\end{equation}
where the potential is sourced by all components of the fluid. In this case, $\rho_k$ and $P_k$ are the total density and pressure of each fluid. We define $\rho_k=\bar{\rho}_k+\delta\rho_k$ and $P_k=\bar{P}_k+\delta P_k$, where the over-barred quantities represent the background.

The background continuity equation for the fluid $i$ is
\begin{equation}
 \dot{\bar{\rho}}_i + 3H\left(\bar{\rho}_i+\frac{\bar{P}_i}{c^2}\right)=0\,,
\end{equation}
where $\bar{\rho}_i=\tfrac{3H^2\Omega_{i}}{8\pi G}$ and $\Omega_{i}$ is the fluid density parameter. To solve the previous expression, it is necessary to specify a relation between pressure and density. This is usually done by introducing the background equation-of-state parameter $w_i=\bar{P}_i/(\bar{\rho}_ic^2)$, so the following equation is solved $\dot{\bar{\rho}}_i + 3H(1 + w_i)\bar{\rho}_i = 0$, once the time dependency of $w_i$ is provided.

To study the perturbations, we introduce comoving coordinates $\vec{x}=\vec{r}/a$, being $a$ the scale factor, and define
\begin{align}
 \rho_i(\vec{x},t) = &\, \bar{\rho}_i(1+\delta_i(\vec{x},t))\,, \label{equ:rpert} \\
 P_i(\vec{x},t) = &\, \bar{P}_i + \delta P_i\,, \label{equ:ppert}\\
 \Phi(\vec{x},t) = &\, {\Phi_0}(\vec{x},t)+\phi(\vec{x},t)\,, \label{equ:fpert}\\
 \vec{v}_i(\vec{x},t) = &\, a[H(a)\vec{x}+\vec{u}_i(\vec{x},t)]\,, \label{equ:vpert}
\end{align}
where $H(a)$ is the Hubble function and $\vec{u}(\vec{x},t)$ the comoving peculiar velocity. We relate pressure perturbations to density perturbations by introducing the effective sound speed $c_{{\rm s},i}^2=\delta P_i/(\delta\rho_ic^2)$. Notice that in a standard cold dark matter model $c_{{\rm s},i}^2=0$, as there are no pressure perturbations.

Inserting Eqs.~(\ref{equ:rpert})--(\ref{equ:vpert}) into Eqs.~(\ref{equ:cnpert})--(\ref{equ:pnpert}), and taking into account the background prescription, we derive the following equations for the perturbed quantities:
\begin{align}
& \dot{\delta}_i + 3H\left(c_{{\rm s},i}^{2}-w_i\right)\delta_i  = 
 -\left[1+w_i+\left(1+c_{{\rm s},i}^{2}\right)\delta_i\right] 
\vec{\nabla}\cdot\vec{u}_i\,, \label{equ:cont-pert2}\\
& \dot{\vec{u}}_i+2H\vec{u}_i+(\vec{u}_i\cdot\vec{\nabla})\vec{u}_i+\frac{\vec{\nabla}\phi}{a^2} = 0\,, \label{equ:euler-pert2}\\
& \nabla^{2}\phi = 4\pi Ga^2\sum_k\bar{\rho}_{k}\left(1+3c_{{\rm s}, k}^{2}\right)\delta_k\,. \label{equ:poisson-pert2}
\end{align}
Note that, as commonly done, we assumed a top-hat profile for the density perturbations. This leads to $\vec{\nabla}\delta_i=0$, which considerably simplifies the equations. 
In addition, both $w_i$ and $c_{{\rm s},i}^{2}$ are functions of time only. Although this is justified for the equation of state, it is a simple approximation for the sound speed case, nevertheless, it is in agreement with the current literature \cite{Kopp2016}.

In order to compute the linear matter power spectrum for GDM, we have followed the prescription given in \cite{Thomas2016,Tutusaus2018}, where the authors modified the Einstein-Boltzmann solver \texttt{CLASS} \citep{Lesgourgues:2011re, Blas:2011rf}.
Thus, at a linear level of perturbations and in the synchronous gauge, the conservation of the energy momentum tensor yields to the equations \citep{1995ApJ...455....7M}
\begin{subequations}
\begin{align}
 & \dot{\delta} + (1+w_{\rm gdm})\left( \theta + \frac{\dot{h}}{2} \right) + 
  3H\left(\frac{\delta P}{\delta\rho} -w_{\rm gdm} \right)\delta = 0 \,,\\
 & \dot{\theta} + H(1-3w_{\rm gdm})\,\theta + \frac{\dot{w}_{\rm gdm}}{1+w_{\rm gdm}}\,\theta -
 \frac{\delta P/\delta\rho}{1+w_{\rm gdm}}\,k^2 \delta + \nonumber\\
 & \qquad k^2\sigma = 0 \,.
\end{align}
\end{subequations}
The system is closed by providing the expressions that relate the GDM equations of state parameter $w_{\rm gdm}$,  the pressure perturbation $\delta P$ and the scalar anisotropic stress $\sigma$ to the density fluctuations $\delta$, the divergence of its velocity $\theta$ and the synchronous metric perturbations $h$ and $\eta$. That is, 
\begin{align}
 & \delta P = c_{\rm s,gdm}^2 \delta\rho - \dot{\rho}\,(c_{\rm s,gdm}^2-c_{\rm a,gdm}^2)\,\theta/k^2\,,\\
 & \dot{\sigma}+3H\frac{c_{\rm a,gdm}^2}{w_{\rm gdm}}\sigma = \frac{4}{3}\frac{c_{\rm vis,gdm}^2}{1+w_{\rm gdm}}(2\theta+\dot{h}+6\dot{\eta}) \label{eq:sigmaEvol}\,
\end{align}
where the adiabatic sound speed is $c_{\rm a,gdm}^2 \equiv (w_{\rm gdm}\,\bar{\rho})\,\dot{}/\dot{\bar{\rho}}$
and $c_{\rm vis,gdm}^2$ is a viscosity parameter that we set to zero.
The public version of CLASS already includes a parameterization of a dark energy fluid with a constant equation of state parameter and constant sound velocity \cite{Lesgourgues:2011re}. We made use of this parameterization to describe GDM, while keeping the cosmological constant for the dark energy contribution

\section{Theoretical predictions for Euclid-like observables}\label{sec:thpred}

As it will be described in next section, the forecasting methods and tools used in this paper are the same of \citet{Blanchard:2019oqi}. However, we must notice here the change of the recipes used to compute theoretical predictions for the \Euclid observables. 

Following \citet{Blanchard:2019oqi} and standard \Euclid forecast practice, the GCsp modelling is based on the linear matter power spectrum, with nonlinear effects entering only through the RSD (FoG) and Alcock–Paczynski terms. In contrast, the nonlinear GDM power spectrum obtained from simulations is employed for the WL, GCph, and XC probes, where nonlinear corrections dominate.

\subsection{Photometric survey}\label{sec:photo}

For the Euclid-like photometric survey, the observables that need to be computed and compared with the data are the angular power spectra for WL, GC$_{\rm ph}$ and their cross correlation, XC.

In \citet{Blanchard:2019oqi} these were calculated, using the Limber approximation, plus the flat-sky approximation with pre-factor set to unity in a flat \lcdm\ Universe, as 
\begin{equation}
 C^{XY}_{ij}(\ell) = \frac{c}{H_0}\int_{z_{\rm min}}^{z_{\rm max}}\de z\,{\frac{W_i^X(z)W_j^Y(z)}{E(z)r^2(z)}P_{\delta\delta}(k_\ell,z)} .\label{eq:ISTrecipe}
\end{equation}
Here, $k_\ell=(\ell+1/2)/r(z)$, $r(z)$ is the comoving distance to redshift $z=1/a-1$, and $P_{\delta\delta}(k_\ell,z)$ corresponds to  the non-linear power spectrum of matter density fluctuations, $\delta$, at wave number $k_{\ell}$ and redshift $z$, when considering the redshift range of the integral from $z_{\rm min}=0.001$ to $z_{\rm max}=4$. 
In addition, the dimensionless Hubble function is defined as $E(z)=H(z)/H_0$. From now on, in all subsequent equations $H_0$ will be  expressed in units of $\mathrm{km/s/Mpc}$.

For each tomographic redshift bin $i$, the window functions $W^X_i(z)$ with $X=\{{\rm L,G}\}$ (corresponding to WL and GCph, respectively) need to be computed differently with respect to what was done in \citet{Blanchard:2019oqi} where fiducial cold dark matter particles were assumed.
We can therefore use the recipe of \Cref{eq:ISTrecipe} accounting for the effects of our $\Lambda$GDM model to calculate $H$, $r$ and $P_{\delta\delta}$, provided by dedicated Boltzmann solvers. The window function for the galaxy clustering can be written as 
\begin{equation}
 W_i^{\rm G}(k,z) =\; (H_0/c)b_i(k,z)\frac{n_i(z)}{\bar{n}_i}E(z)\,, \label{eq:wg_mg}
 \end{equation}
 while the window function for lensing becomes 

 \begin{align}
 W_i^{\rm L}(k,z) =&\; (H_0/c)^2\frac{3}{2}\Omega_{\rm m,0}\, (1+3\,c^2_{s,{\rm gdm}}) \, (1+z)^{(1+3\,w_{\rm gdm})}\,r(z)\\&\Sigma(k,z)
\int_z^{z_{\rm max}}{\de z'\frac{n_i(z)}{\bar{n}_i}\frac{r(z'-z)}{r(z')}}\nonumber\\
  &+W^{\rm IA}_i(k,z)\, , \label{eq:wl_mg}
\end{align}
where $n_i(z)/\bar{n}_i$ and $b_i(k,z)$ are the normalised galaxy distribution and the galaxy bias in the $i$-th redshift bin, respectively; while $W^{\rm IA}_i(k,z)$ encodes the contribution of intrinsic alignments (IA) to the WL power spectrum. We follow \citet{Blanchard:2019oqi} when assuming an effective scale-independent linear galaxy bias. 

We adopted this choice to enable a direct comparison with the standard analysis, keeping the GDM extension as the only varying element. Introducing a scale-dependent galaxy bias would add extra degrees of freedom, making it harder to distinguish the effects of the GDM model from those of the bias model. 

The IA contribution is computed following the eNLA model from \citet{Blanchard:2019oqi}. In our case, this term becomes
\begin{equation}\label{eq:IA}
 W^{\rm IA}_i(k,z)=-\frac{\mathcal{A}_{\rm IA}\mathcal{C}_{\rm IA}\Omega_{\rm m,0}\,(1+3\,c^2_{s,{\rm gdm}})\,\mathcal{F}_{\rm IA}(z)}{\delta(k,z)/\delta(k,z=0)}\frac{n_i(z)}{\bar{n}_i(z)}(H_0/c)E(z)\,,
\end{equation}
where 
\begin{equation}
 \mathcal{F}_{\rm IA}(z)=(1+z)^{\eta_{\rm IA}}\left[\frac{\langle L\rangle(z)}{L_\star(z)}\right]^{\beta_{\rm IA}}\,.
\end{equation}
Here, $\langle L\rangle(z)$ and $L_\star(z)$ denote the redshift-dependent mean and characteristic luminosities of the source galaxies, respectively, as derived from the luminosity function. The parameters $\mathcal{A}{\rm IA}$, $\beta{\rm IA}$, and $\eta_{\rm IA}$ are the nuisance parameters of the intrinsic-alignment model, while $\mathcal{C}_{\rm IA}$ is a constant introduced to ensure dimensional consistency.

Deviations from the standard CDM properties affect the IA contribution through the modified growth of matter perturbations. This effect is explicitly accounted for in \Cref{eq:IA} via the matter over density $\delta(k,z)$.

In our model, the background evolution differs from $\Lambda$CDM, which leads to modifications in the expansion rate $E(z)$ and the comoving distance $r(z)$ appearing \Cref{eq:wl_mg}. Accordingly, the standard prefactor $(H_0/c)^2\,\tfrac{3}{2}\,\Omega_m(1+z)$—valid for a cold dark matter scenario neglecting baryons and, a fortiori, relativistic contributions to the lensing potential—must be adjusted. Since the $\Lambda$GDM framework does not modify the Weyl potential, the relation between $(\Phi+\Psi)$ and the matter overdensity is the standard one; in our notation this corresponds to setting, when $X$ and $Y$ refer to lensing, $\Sigma=1$ in \Cref{eq:ISTrecipe}. Therefore, the geometric part of the lensing kernel entering \Cref{eq:wl_mg} reduces to its standard $\Lambda$CDM form.

\subsection{Non-linear modelling}\label{sec:non-linear}
As no general analytical solution exists for the deeply non-linear power spectrum in scenarios where dark matter properties differ from those of CDM, we performed cosmological simulations to compute this quantity within the GDM framework.
To this end, we set the initial conditions for the GDM model by generating an initial realization of $N$-body particles derived from the distribution of density perturbations specified by a given linear matter power spectrum. We then evolved these particles using an $N$-body solver. The simulations were performed under the approximation that small-scale particle collisions are negligible over time. In other words, we assumed that the differences between the $\Lambda$GDM and $\Lambda$CDM models arise solely from the initial conditions, that means neglecting the GDM thermal velocities. A similar approach was adopted in \citet{10.1093/mnras/stac1925}.

This approximation is justified because the expected velocity dispersion is of the same order as the sound speed adopted in this work, $c_{\rm s} \sim 10^{-3}$ km $\mathrm{s^{-1}}$\citep{Kunz2016}. Such a value is comparable to those typically used in warm dark matter (WDM) studies, and several orders of magnitude smaller than the characteristic velocities adopted in cosmological simulations that include massive neutrinos ($\sim 100$ km $\mathrm{s^{-1}}$). Consequently, GDM particles can effectively be treated as cold, since their velocities are lower than those of the cosmic microwave background (CMB) photons and have a negligible impact on structure formation. Moreover, at low redshifts, the peculiar velocities induced by gravitational clustering rapidly dominate over any initial thermal motion, further supporting this approximation.

Our assumption is supported by similar studies, such as the one performed for WDM in \citet{Euclid:2024pwi}, which can be considered as a particular case of the GDM framework. In that work, the authors tested the impact of neglecting particle velocities by running an additional realization that explicitly included them in the model. When comparing the residuals between the main \Euclid\ probes—namely, the angular power spectra of galaxy lensing, the cross-correlation between galaxy clustering and lensing, and galaxy clustering alone—computed using the nonlinear power spectra of both approaches, they found differences below the percent level. These deviations are well within the expected sample variance \Euclid \ of $\sim 1.6\%$, leading to the conclusion that their assumption, similar to ours, does not introduce significant systematic effects in the forecast analysis.

\begin{figure}[htbp]
    \centering
    \includegraphics[width=0.95\linewidth]{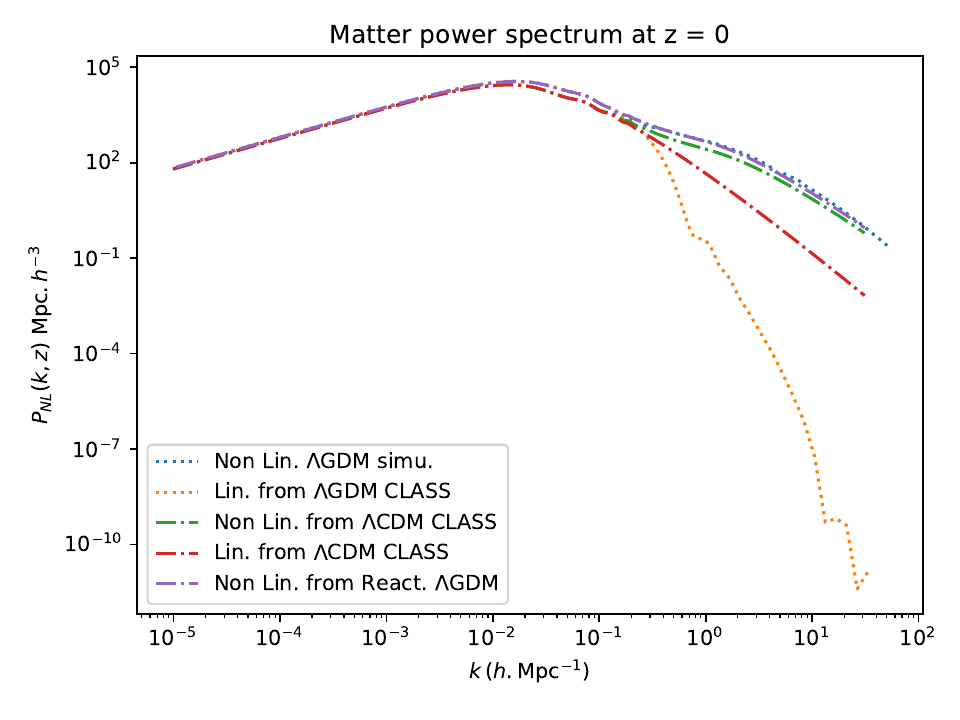}
    \caption{Showing broad agreement of the GDM model matter power spectrum at linear and nonlinear scales between ones obtained from GDM simulations versus ones following the response approximate method \cite{Mead:2016ybv} }
    \label{fig:responsevalidation}
\end{figure}

Finally, the power spectrum obtained from our simulations agrees, within $\sim10\%$, with the nonlinear power spectrum derived using the response approximation method of \citet{Mead:2016ybv}, as shown in \Cref{fig:responsevalidation}. In this approach, the nonlinear boost is determined by rescaling the $\Lambda$CDM power spectrum using the ratio of the $\Lambda$CDM one–halo term to that computed with the spherical–collapse parameters, $\delta_{\rm c}$ and $\Delta_{\rm vir}$, evaluated within the $\Lambda$GDM model as in \citet{Pace:2019vrs}.

\begin{figure}
	\centering
	\includegraphics[width=\linewidth]{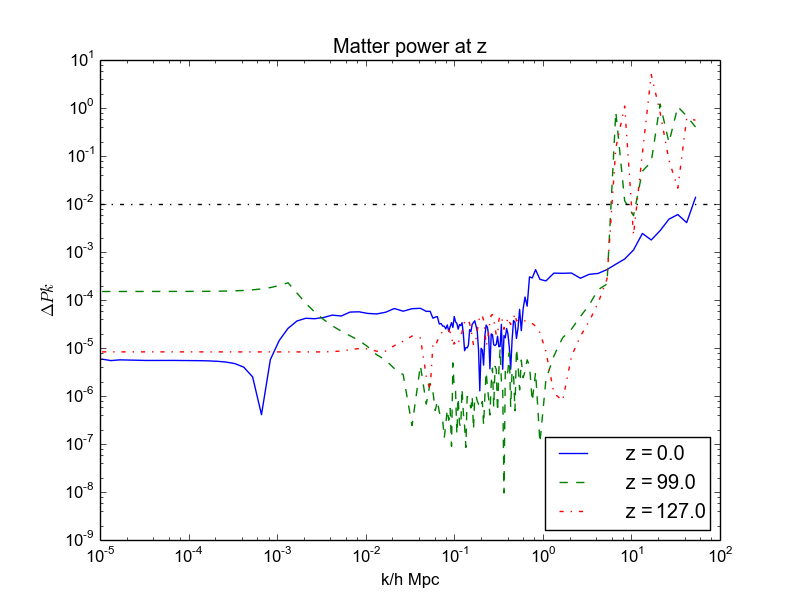}
	\caption{   
    Relative differences between the linear matter power spectrum computed with the \texttt{GDM-class} code and our private version, shown at redshifts $z = 0$, $z = 99$, and $z = 127$.}
	\label{fig:linear_pk}
\end{figure}

\begin{figure}
	\centering
	\includegraphics[width=0.95\linewidth]{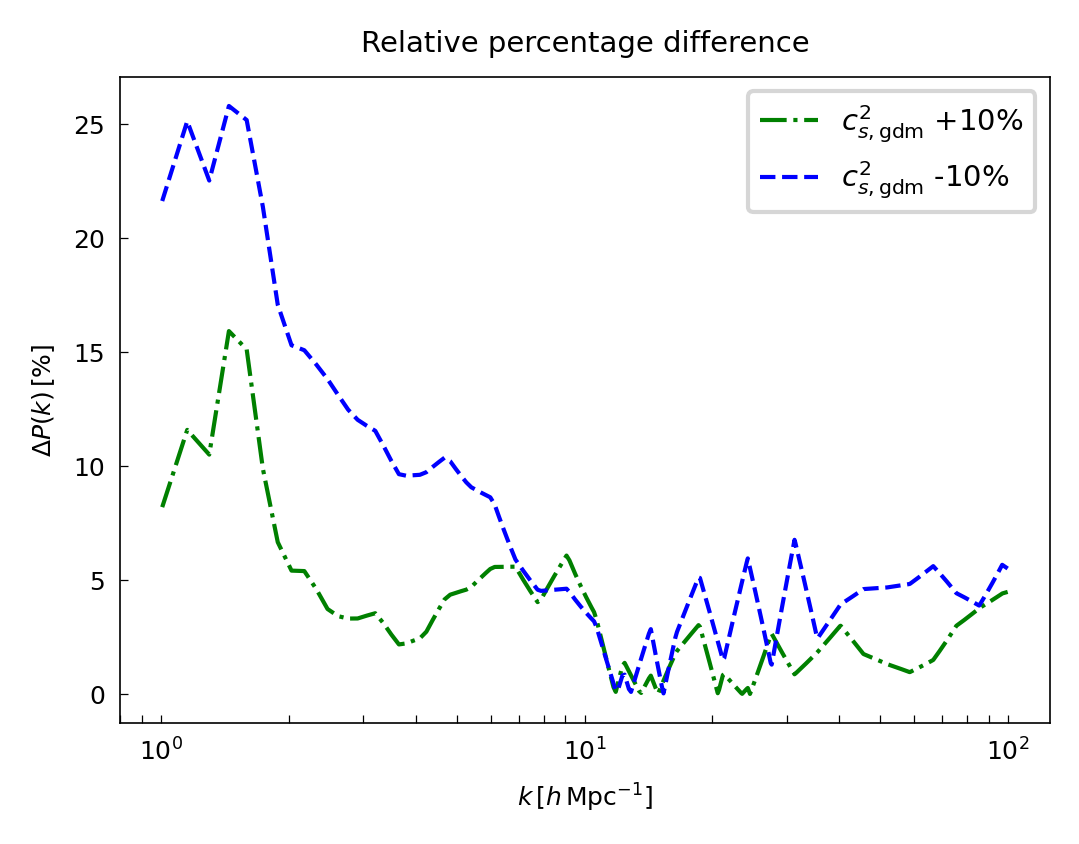}
	\caption{Relative percent difference in the nonlinear matter power spectrum for $c_{s,{\rm gdm}}^{2}$ varied by $\pm 10\%$ around its fiducial value. 
The impact is most pronounced at intermediate nonlinear scales, where deviations are maximised.
}
	\label{fig: nonlinear_csg}
\end{figure}

To generate the initial conditions, we computed the linear matter power spectrum using the \texttt{GDM-class} code \citep{ilic2021dark}, which we validated against our private version of the code. The comparison between the two codes at different redshifts is shown in Fig.~\ref{fig:linear_pk}.
We observe that the differences remain at the $\lesssim 10^{-4}$ level over the relevant range of scales. We used the following parameter values: $w_{\text{gdm}} = 1\times10^{-4}$ and $c_{\text{s,gdm}}^2 = 1.1\times10^{-6}$ and based on this spectrum, the initial density field was generated using the \texttt{N-GenIC} code \citep{Springel2005,angulo2012scaling}. The subsequent evolution of the $N$-body particles was followed the TreePM N-body code \texttt{Gadget-4} \citep{Springel2005}, which employs a hierarchical tree algorithm to compute gravitational forces. This setup allowed us to obtain the particle distributions at different redshifts and to compute the corresponding nonlinear matter power spectra for each GDM scenario. As an illustrative example, Figure~\ref{fig: nonlinear_csg} shows the relative percent difference in the nonlinear matter power spectrum when varying $c_{s,\rm gdm}^2$ by $\pm 10\%$ around its fiducial value. As observed, the largest deviations occur at intermediate nonlinear scales, where the sensitivity of the power spectrum to $c_{s,\rm gdm}^2$ is maximal while for large scales the percent difference is significantly smaller.

In all our simulations, the box size was set to $L = 100,h^{-1}\mathrm{Mpc}$, with a total particle number of $N_{\text{tot}} = 512^3$ and an initial redshift of $z_{\text{ini}} = 127$.
The adopted cosmological parameters are $\Omega_m = 0.268$, $\Omega_b = 0.05$, $H_0 = 67,\mathrm{km,s^{-1},Mpc^{-1}}$ and $n_s=0.96$. We employed a shooting method to iteratively adjust the primordial amplitude $A_{\rm s}$ until the resulting linear power spectrum matched the target value of $\sigma_8$ at $z=0$. 
We considered three different scenarios: $\sigma_8=0.81$ (GDM I), $\sigma_8=0.67$ (GDM II) and $\sigma_8=0.77$ (GDM III).
The first and third choices for $\sigma_8$ are motivated by the objective of testing the ability of \Euclid survey to distinguish between the amplitude of matter fluctuation at low and high redshifts. The second value, instead, corresponds to the $\sigma_8$ obtained when fixing the primordial amplitude of scalar perturbations, $A_{\rm s}$, to the value constrained by CMB observations, allowing us to assess to what extent local data can rule out the $\Lambda$GDM model under this normalization.

Note that our fiducial cosmology includes massive neutrinos with a total mass of $\sum m_\nu=0.06\,\mathrm{eV}$, which we keep fixed in the Fisher matrix analysis (see section \Cref{sec:fisher}).

\subsection{Spectroscopic survey}\label{sec:spect}
In order to exploit data from our supposed Euclid-like spectroscopic survey, we need to compute the theoretical prediction for the observed galaxy power spectrum in the extended model considered here. The full non-linear model for the observed galaxy power spectrum is given by 
\begin{multline}
P_\text{obs}(k_\text{ref},\mu_{\theta,\text{ref}};z) = 
\frac{1}{q_\perp^2(z) q_\parallel(z)} 
\left\{\frac{\left[b\sigma_8(k,z)+f\sigma_8(k,z)\mu_{\theta}^2\right]^2}{1+k^2\mu_{\theta}^2\sigma_{\rm p}^2(z)}\right\} 
\\
\times \frac{P_\text{dw}(k,\mu_{\theta};z)}{\sigma_8^2(z)}  
F_z(k,\mu_{\theta};z) 
+ P_\text{s}(z) \, , 
\label{eq:GC:pk-ext}
\end{multline}
here the $P_{\rm dw}(k,\mu;\,z)$ is the de-wiggled power spectrum which
models the smearing of the BAO features due to the displacement field of wavelengths smaller than the BAO scale,
\begin{equation}
P_\text{dw}(k,\mu;z) = P_{\delta\delta}(k;z)\,\text{e}^{-g_\mu k^2} + P_\text{nw}(k;z)\left(1-\text{e}^{-g_\mu k^2}\right) \, .
\label{eq:GC:pk_dw}
\end{equation}

The $P_{\rm nw}(k;z) $ is a ‘no-wiggle’ power spectrum with the same broad band shape as $P_{\delta\delta}(k;z) $ but without BAO features. The term in brackets in \Cref{eq:GC:pk-ext} is the RSD contribution corrected for the non-linear finger-of-God (FoG) effect, where we defined $b\sigma_8(k,z)$ as the product of the effective scale-dependent bias of galaxy samples and the r.m.s.\ matter density fluctuation $\sigma_8(z)$; similarly, $f\sigma_8(k,z)$ is the product of the scale-dependent growth rate and $\sigma_8(z)$; and, finally,  $\mu_{\theta}$ is the cosine of the angle $\theta$ between the wave vector $\bm k$ and the line-of-sight direction $\hat{\bm r}$.

The observed galaxy power spectrum is modulated by the redshift uncertainties which are manifested as a smearing of the galaxy density field along the line-of-sight, hence the factor $F_z$ in \Cref{eq:GC:pk-ext} reads 
\begin{equation}
    F_z(k, \mu_{\theta};z) = \text{e}^{-k^2\mu_{\theta}^2\sigma_{r}^2(z)}\,,
\end{equation}
being $\sigma_{r}^2(z) = c(1+z)\sigma_{0,z}/H(z)$ and $\sigma_{0,z}$ is the error on the measured redshifts. 

The Alcock-Paczynski effect is parameterised in terms of the angular diameter distance $D_{\rm A}(z)$ and the Hubble parameter $H(z)$ as 
\begin{align}
q_{\perp}(z) &= \frac{D_{\rm A}(z)}{D_{\rm A,\, ref}(z)},\\
q_{\parallel}(z) &= \frac{H_\text{ref}(z)}{H(z)}\,.
\end{align}

Finally, the $P_{\rm s}(z)$ is a scale-independent shot noise term, which enters as a nuisance parameter \citepalias[see][]{Blanchard:2019oqi}.

The two phenomenological parameters related to the velocity dispersion, $\sigma_{\rm v}$, and the pairwise velocity dispersion, $\sigma_{\rm p}$ are given by
\begin{align}
\sigma^2_{\rm v}(z, \mu_{\theta}) &= \frac{1}{6\pi^2}\int\de k\, P_{\delta\delta}(k,z)\left\{1 - \mu_{\theta}^2 + \mu_{\theta}^2\left[1+f(k,z)\right]^2\right\},\label{eq:sigmav}\\
\sigma_{\rm p}^2(z) &= \frac{1}{6\pi^2}\int\de k\, P_{\delta\delta}(k,z)f^2(k,z)\,.\label{eq:sigmap}
\end{align}

These parameters account for the damping of the BAO features and the FoG effect, respectively. The smearing of the BAO peak is due to the bulk motion of scales smaller than the BAO scale. For the power spectrum, this can be modeled in the Zeldovich approximation by a multiplicative damping term of the form 
$\exp\left[-k^i k^j \langle d^i (z) d^j(z) \rangle\right]$,
where $\langle d^i (z) d^j(z) \rangle$ is the correlation function of the displacement field $d^i$ evaluated at zero distance (see for instance the appendix~C of \citealt{Peloso:2015jua} for more details). Finally, we would like to clarify that in \Cref{eq:GC:pk_dw} we used the function $g_{\mu}$ to express the damping of the BAO features in the matter power spectrum to keep the recipe closer to the \citet{Blanchard:2019oqi}; in this work, $g_\mu = \sigma_{\rm v}(z,\,\mu_{\theta})$.

Finally, due to the scale dependence of $\sigma_{\rm p}$ and $\sigma_{\rm v}$, we evaluated both parameters at each redshift bin at the fiducial and let them free in the Fisher matrix analysis. 

\section{Survey specifications and analysis method}\label{sec:fisher}
In order to forecast constraints on this model, we will consider the GDM I, GDM II and GDM III  scenarios \Cref{sec:non-linear}. The fiducial cosmological parameter vectors read as follows
\begin{align}
    &\bm\Theta \phantom{_{\rm fid,1}}=\{\Omega_{\rm m,0},\, \Omega_{\rm b,0},\, h,\, n_{\rm s},\, \sigma_8,\, \ w_{GDM}\ \ {c^2_s}_{gdm}\}\,,\nonumber\\ 
    &\bm\Theta_{\rm fid,GDM \, I}=\{ 0.268,\, 0.05,\, 0.67,\, 0.96,\, 0.81,\, 10^{-4},\ 5\times 10^{-6}\}\,,\nonumber \,\\
    &\bm\Theta_{\rm fid,GDM \, II}=\{ 0.268,\, 0.05,\, 0.67,\, 0.96,\, 0.67,\, 10^{-4},\ 5\times 10^{-6}\}\,,\nonumber \,\\
    &\bm\Theta_{\rm fid,GDM \, III}=\{ 0.268,\, 0.05,\, 0.67,\, 0.96,\, 0.77,\, 10^{-4},\ 5\times 10^{-6}\}\,.
    \label{eq:fiducial-params}
\end{align}

Concerning the photometric probes, the galaxy distribution is binned into $10$ equi-populated redshift bins with an overall distribution following
\begin{equation}
    n(z)\propto\left(\frac{z}{z_0}\right)^2\,\text{exp}\left[-\left(\frac{z}{z_0}\right)^{3/2}\right]\,,
\end{equation}
with $z_0=0.9/\sqrt{2}$ and the normalisation set by the requirement that the surface density of galaxies is $\bar{n}_g=30\,\mathrm{arcmin}^{-2}$. The redshift distribution is then convolved with a sum of two Gaussian distributions to account for the photometric redshift uncertainties \citepalias[see][for details]{Blanchard:2019oqi}. The galaxy bias is assumed to be constant within each redshift bin, and its values $b_i$ are introduced as nuisance parameters in our analysis. Their fiducial values are determined by $\sqrt{1+\bar{z}_i}$, where $\bar{z}_i$ is the mean redshift of each redshift bin. 

Moreover, we follow \citet{Blanchard:2019oqi} in accounting for a Gaussian covariance between the different photometric probes:
\begin{multline}
    \text{Cov}\left[C_{ij}^{AB}(\ell),C_{kl}^{CD}(\ell')\right]=\frac{\delta_{\ell\ell'}^{\rm K}}{(2\ell+1)f_{\rm sky}\Delta \ell}\\
    \times\left\{\left[C_{ik}^{AC}(\ell)+N_{ik}^{AC}(\ell)\right]\left[C_{jl}^{BD}(\ell')+N_{jl}^{BD}(\ell')\right]\right.\\
    +\left.\left[C_{il}^{AD}(\ell)+N_{il}^{AD}(\ell)\right]\left[C_{jk}^{BC}(\ell')+N_{jk}^{BC}(\ell')\right]\right\}\,,
\end{multline}
where upper-case Latin indexes $A,\ldots=\{{\rm WL},\,GCph\}$, lower-case Latin indexes $i,\ldots$ run over all tomographic bins, $\delta_{\ell\ell'}^{\rm K}$ is the Kronecker delta symbol, $f_{\rm sky}\simeq0.36$ represents the fraction of the sky observed by \Euclid, and $\Delta \ell$ denotes the width of the multipole bins, where we use $100$ equi-spaced bins in log-space. The noise terms, which for the observables considered here are in fact white noise, namely $N_{ij}^{AB}(\ell)\equiv N_{ij}^{AB}$, read
\begin{align}
    N_{ij}^{\rm LL}(\ell) &= \frac{\delta_{ij}^{\rm K}}{\bar{n}_i}\sigma_\epsilon^2\,,\\
    N_{ij}^{\rm GG}(\ell) &= \frac{\delta_{ij}^{\rm K}}{\bar{n}_i}\,,\\
    N_{ij}^{\rm GL}(\ell) &= 0\,,
\end{align}
where $\sigma_\epsilon^2=0.3^2$ is the variance of observed ellipticites.

For the spectroscopic probe, we evaluate the Fisher matrix for the 
observed galaxy power spectrum according to the recipe outlined in \citet{Blanchard:2019oqi}, where, for a parameter vector $p_\alpha$, $F_{\alpha\beta}(z_i)$ is given by 
\begin{equation}
F_{\alpha\beta}^{\text{GC}}(z_i)=\frac{1}{8\pi^{2}}\int_{-1}^{1}{\mathrm{d}u}\int_{k_{\text{min}}}^{k_{\text{max}}}k^{2}V_{\text{eff}}\frac{d\log P}{dp_{\alpha}}\biggl|_{f}\frac{d\log P}{dp_{\beta}}\biggl|_{f}\,{\mathrm{d}k}\,,\label{eq:fmgc}
\end{equation}
where the effective survey volume, $V_{\text{eff}}$, is
\begin{equation}
V_{\text{eff}}(k,\mu; z)=\left(\frac{\bar n_{{\rm gal},i}(z)P_{\rm obs}(k,u;z)}{\bar n_{{\rm gal},i}(z)P_{\rm obs}(k,u;z)+1}\right)^{2}V_{\text{i}}(z)\label{veff} \,,
\end{equation}
and $V_i$ is the redshift bin volume, $\bar n_{{\rm gal},i}$ the galaxy number density in each bin and $P_{\rm obs}$ calculated at the fiducial.
Here, $\alpha$ and $\beta$ run over the cosmological parameters of the set $\bm\Theta$, the index $i$ labels the redshift bin, each respectively centred in $z_i = \{1.0,\,1.2,\,1.4,\,1.65\}$, whose widths are $\Delta z = 0.2$ for the first three bins and $\Delta z = 0.3$ for the last bin. In this paper, we vary the observed galaxy power spectrum with respect to the cosmological parameters of $\bm\Theta$ directly, plus the additional redshift-dependent parameters $\ln b\sigma_8(z_i)$ and $P_{\rm s}(z_i)$ and the non-linear parameters $\sigma_p$ and $\sigma_v$ that we marginalise over. We consider the numerical values for the galaxy bias, $b(z)$, and the expected number density of the observed H$\alpha$ emitters, $n(z)$,  reported in table 3 of \citet{Blanchard:2019oqi}.

For both probes, we consider two different scenarios: an optimistic and a pessimistic case. In the optimistic case, we consider $k_{\rm max}=0.30\,h\,\mathrm{Mpc}^{-1}$ for GCsp, $\ell_{\rm max}=5000$ for WL, and $\ell_{\rm max}=3000$ for GC$_{\rm ph}$ and XC. Instead, in the pessimistic scenario, we consider $k_{\rm max}=0.25\,h\,\mathrm{Mpc}^{1}$ for GCsp, $\ell_{\rm max}=1500$ for WL, and $\ell_{\rm max}=750$ for GC$_{\rm ph}$ and XC. We also perform a second pessimistic forecast for GCsp\ only, where we set our maximum wave number at $k_{\rm max} = 0.15\,h\,{\rm Mpc}^{-1}$ in order to have a more conservative estimate of the constraining power of the GCsp\ probe.

The reason for this is that the underlying matter power spectrum of \Cref{eq:GC:pk-ext} that we are using in our observed galaxy power spectrum recipe is a linear, as we detailed in \Cref{sec:spect}.
        It is known that non-linear corrections start playing a role above scales of around $k = 0.1\,h\,{\rm Mpc}^{-1}$ for the redshifts under consideration \citep[see][]{Taruya:2010mx} and, therefore, the use of a linear power spectrum beyond these scales can bias our constraints. Hence, we aim to estimate what would happen if we used just quasi-linear scales in the analysis. As a reference for the reader, we list the specific choices of scales and settings used for each observable in \Cref{tab:specifications-ec-survey}.

\begin{table}
	\centering
	\caption{Euclid-like survey specifications for WL, GC$_{\rm ph}$ and GC$_{\rm sp}$. }
	\label{tab:specifications-ec-survey}
	\begin{tabularx}{\columnwidth}{Xll}
		\hline 
		Survey area & $A_{\rm survey}$  & $15\,000\,\deg^2$  \\
		\hline
		\hline
		\multicolumn{3}{c}{WL}\\
		\hline
		Number of photo-$z$ bins & $N_z$ & 10 \\
		Galaxy number density & $\bar n_{\rm gal}$  & $30\,\mathrm{arcmin}^{-2}$ \\
		Intrinsic ellipticity $\sigma$ & $\sigma_\epsilon$  & 0.30 \\
		Minimum multipole & $\ell_{\rm min}$ & 10\\
		Maximum multipole & $\ell_{\rm max}$ & \\
		-- Pessimistic & & $1500$\\
		-- Optimistic & & $5000$\\
        \hline
		\hline
        \multicolumn{3}{c}{GC$_{\rm ph}$}\\
        \hline
		Number of photo-$z$ bins & $N_z$ & 10 \\
		Galaxy number density & $\bar n_{\rm gal}$  & $30\,\mathrm{arcmin}^{-2}$ \\
		Minimum multipole & $\ell_{\rm min}$ & 10 \\
		Maximum multipole & $\ell_{\rm max}$ & \\
		-- Pessimistic & & $750$\\
		-- Optimistic & & $3000$\\
		\hline
		\hline
		\multicolumn{3}{c}{GC$_{\rm sp}$}\\
		\hline
		Number of spectro-$z$ bins & $n_z$ & 4 \\
		Centres of the bins &$z_i$ & $\{1.0,\,1.2,\,1.4,\,1.65\}$\\
		Error on redshift & $\sigma_{0,z}$ & 0.001 \\
		Minimum scale & $k_{\rm min}$ & $0.001\,h\,{\rm Mpc}^{-1}$\\
		Maximum scale & $k_{\rm max}$ & \\
		-- Quasi-linear & & $0.15\,h\,{\rm Mpc}^{-1}$\\
		-- Pessimistic & & $0.25\,h\,{\rm Mpc}^{-1}$\\
		-- Optimistic & & $0.30\,h\,{\rm Mpc}^{-1}$\\
		\hline 
	\end{tabularx}
\end{table}

In this work, as in \citet{Blanchard:2019oqi}, we show the results for most of these single probes, but also for their combinations. It is important to mention that when we consider the combination of GC$_{\rm ph}$ with WL, we neglect any cross-correlation. However, when we add their cross-correlation XC, we include it both in the data vector and in the covariance, i.e.\ we perform a full analysis taking into account the cross-covariances between GC$_{\rm ph}$, WL, and their cross-correlation. In the optimistic scenario, we assume GCsp\ to be uncorrelated to photometric probes. Moreover, the covariance between the photometric and spectroscopic probes was also found by recent studies \cite{taylor2022covariance, Euclid:2024pwi} to be negligible \footnote{These studies were conducted, however, within $\Lambda$CDM models and it might be that some effects will appear when adopting our model, but we leave such kind of investigations to future works.}. A fortiori, in the pessimistic setting, we neglect as well the correlations between GCsp\ and WL.

\section{Results} \label{sec:results}

As discussed in \Cref{sec:non-linear,sec:fisher} and motivated by the observational constraints and existing tensions outlined in \Cref{sec:intro}, we consider two fiducial configurations of the parameters of the $\Lambda$GDM model: one far from the General relativity limit (GR) (GDM I) and one close to it (GDM II). In addition, we define a third fiducial model, GDM III, which corresponds to a “tension-like” value that reflects the level of discrepancy between local and CMB measurements. 
We remind the reader of the two considered scenarios, pessimistic and optimistic, as explained in \Cref{sec:fisher} plus the `quasi-linear' even more pessimistic scenario for GCsp, defined by $k_{\rm max} = 0.15\,h\,{\rm Mpc}^{-1}$. 

\begin{figure*}[ht!]
    \centering
    \begin{tabular}{cc}
    \includegraphics[width=0.5\linewidth]{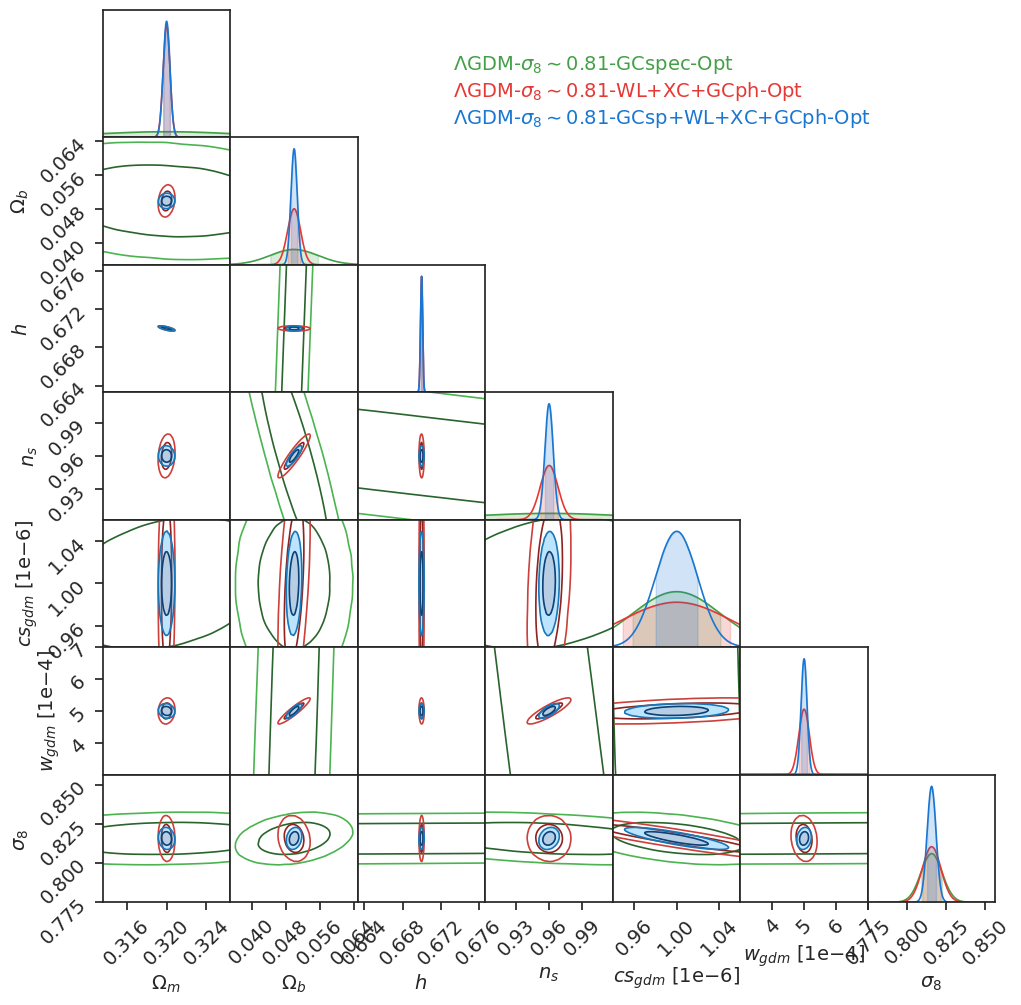}&
    \includegraphics[width=0.5\linewidth]{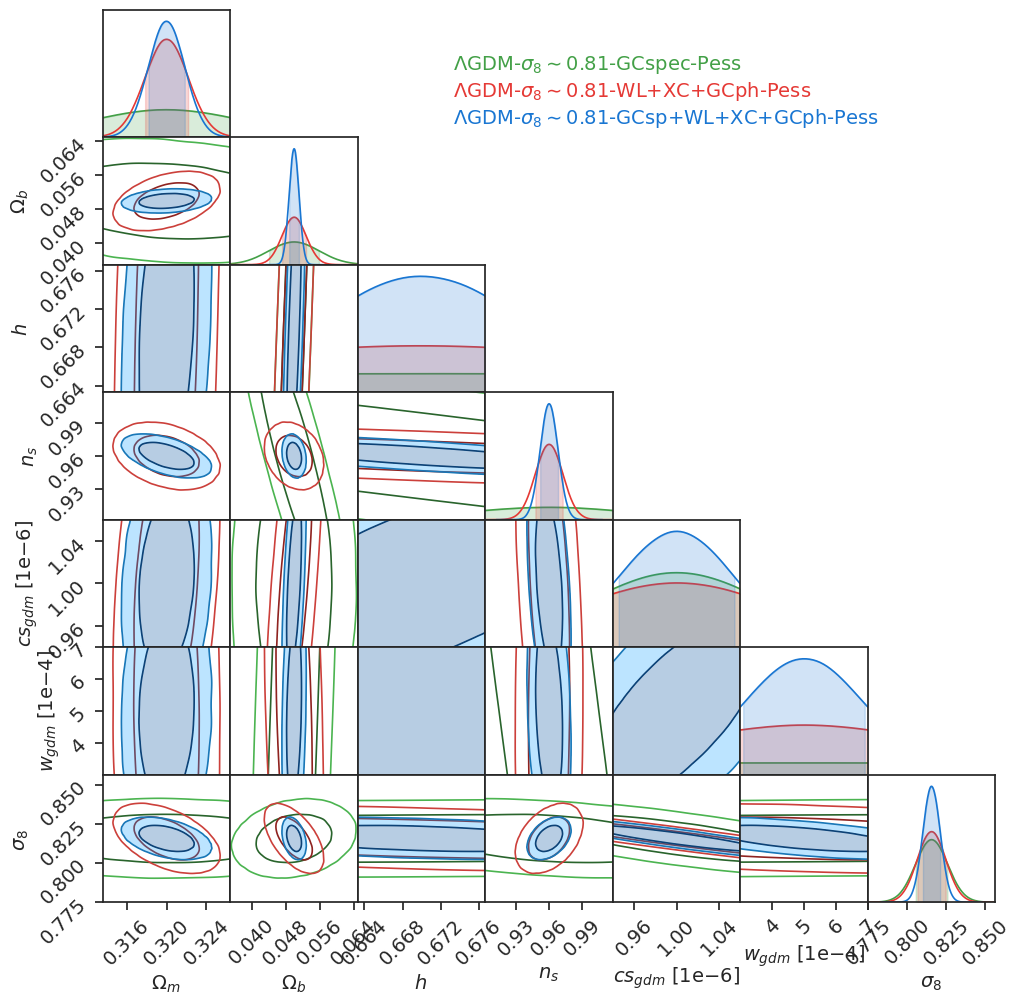}
    \end{tabular}
    \caption{\textbf{Left:} 1 and 2$\sigma$ joint marginal error contours on the cosmological parameters for GDM I model from Fisher forecasts in the optimistic settings. In green using spectroscopic Galaxy Clustering (GCsp), in red the photometric probes and their cross-correlations (WL+GCph+XC) and in blue all the photometric probes with their cross correlation combined with spectroscopic Galaxy Clustering (GC$_{\rm sp}$+WL+GCph+XC).	
	  \textbf{Right:} 1 and 2$\sigma$ joint marginal error contours on the cosmological parameters for the same model with the same previous probe combinations but in the pessimistic settings.}
    \label{fig:081XCGC}
\end{figure*}

\begin{figure*}[ht!]
    \centering
    \begin{tabular}{cc}
    \includegraphics[width=0.5\linewidth]{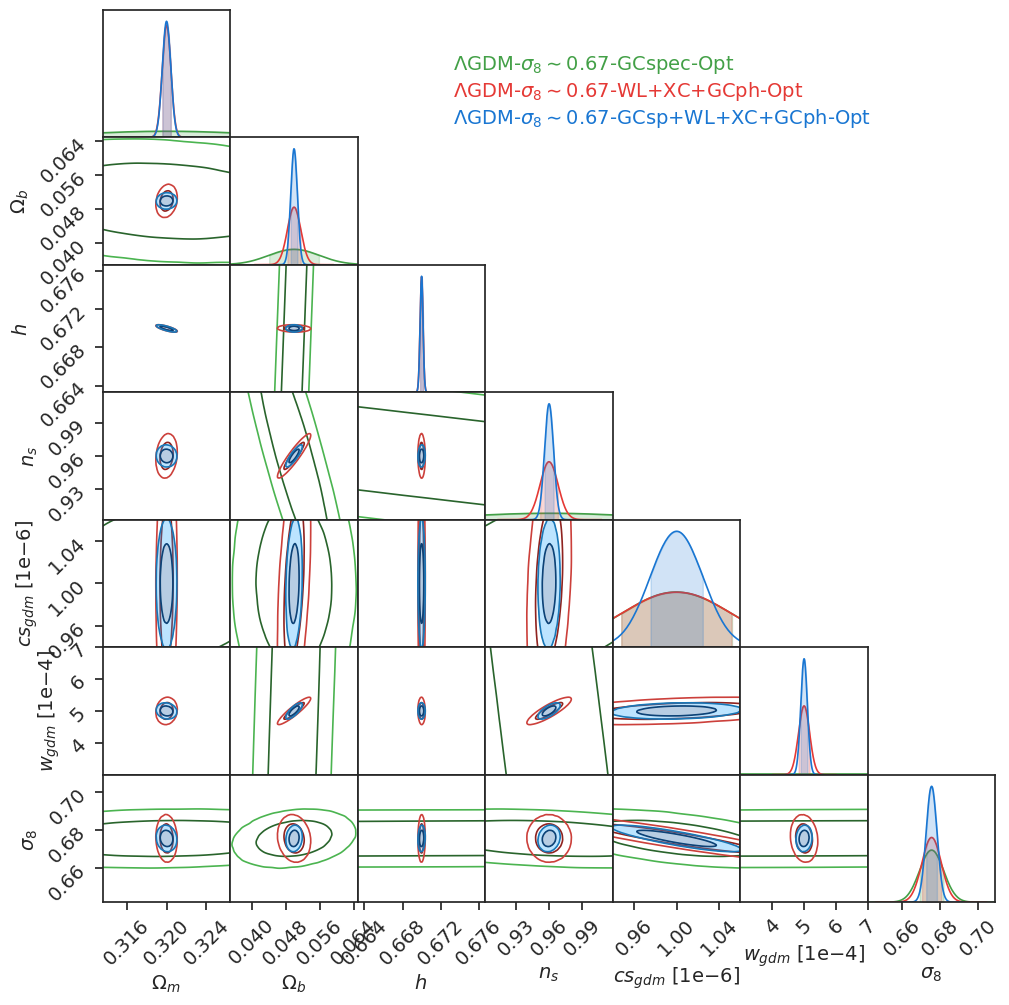}&
    \includegraphics[width=0.5\linewidth]{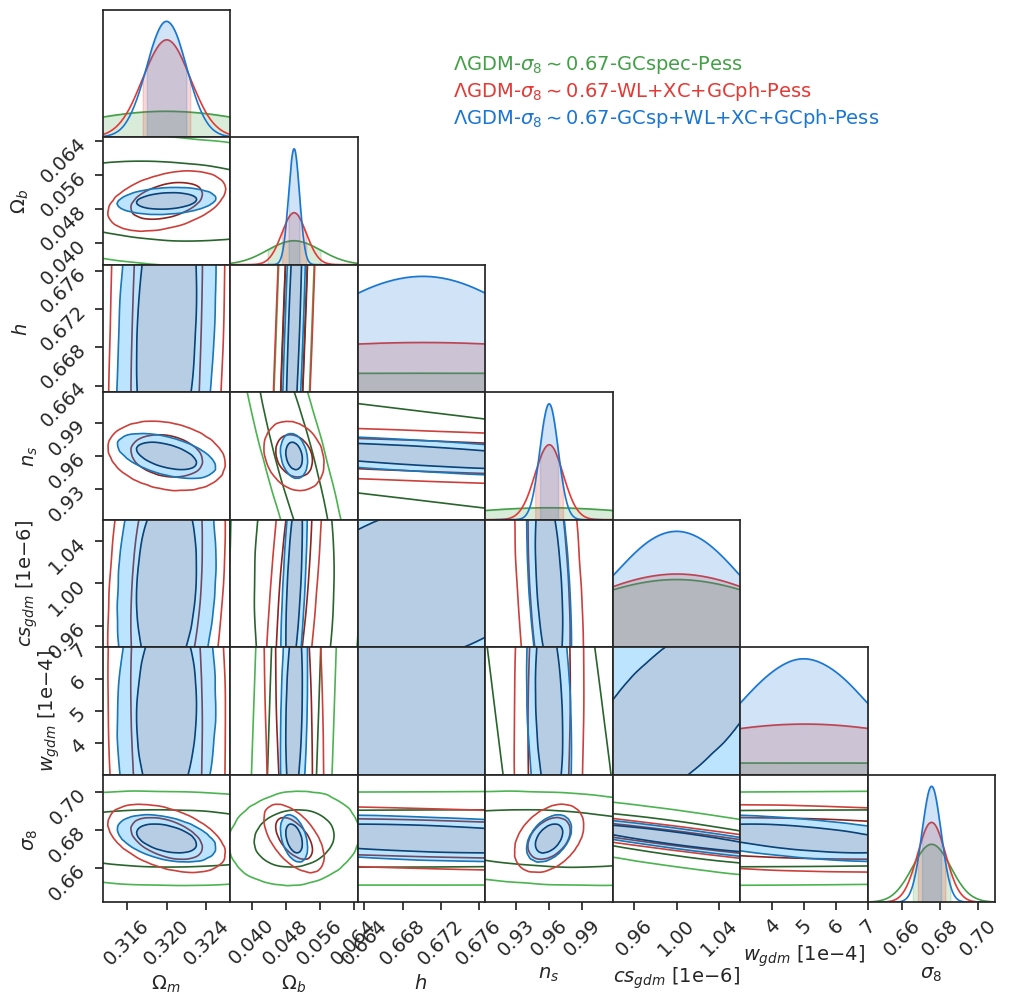}
    \end{tabular}
    \caption{\textbf{Left:} 1 and 2$\sigma$ joint marginal error contours on the cosmological parameters for GDM II model from Fisher forecasts in the optimistic settings. In green using spectroscopic Galaxy Clustering (GCsp), in red the photometric probes and their cross-correlations (WL+GCph+XC) and in blue all the photometric probes with their cross correlation combined with spectroscopic Galaxy Clustering (GC$_{\rm sp}$+WL+GCph+XC).	
	  \textbf{Right:} 1 and 2$\sigma$ joint marginal error contours on the cosmological parameters for the same model with the same previous probe combinations but in the pessimistic settings.}
    \label{fig:067XCGC}
\end{figure*}

\begin{figure*}[ht!]
    \centering
    \begin{tabular}{cc}
    \includegraphics[width=0.5\linewidth]{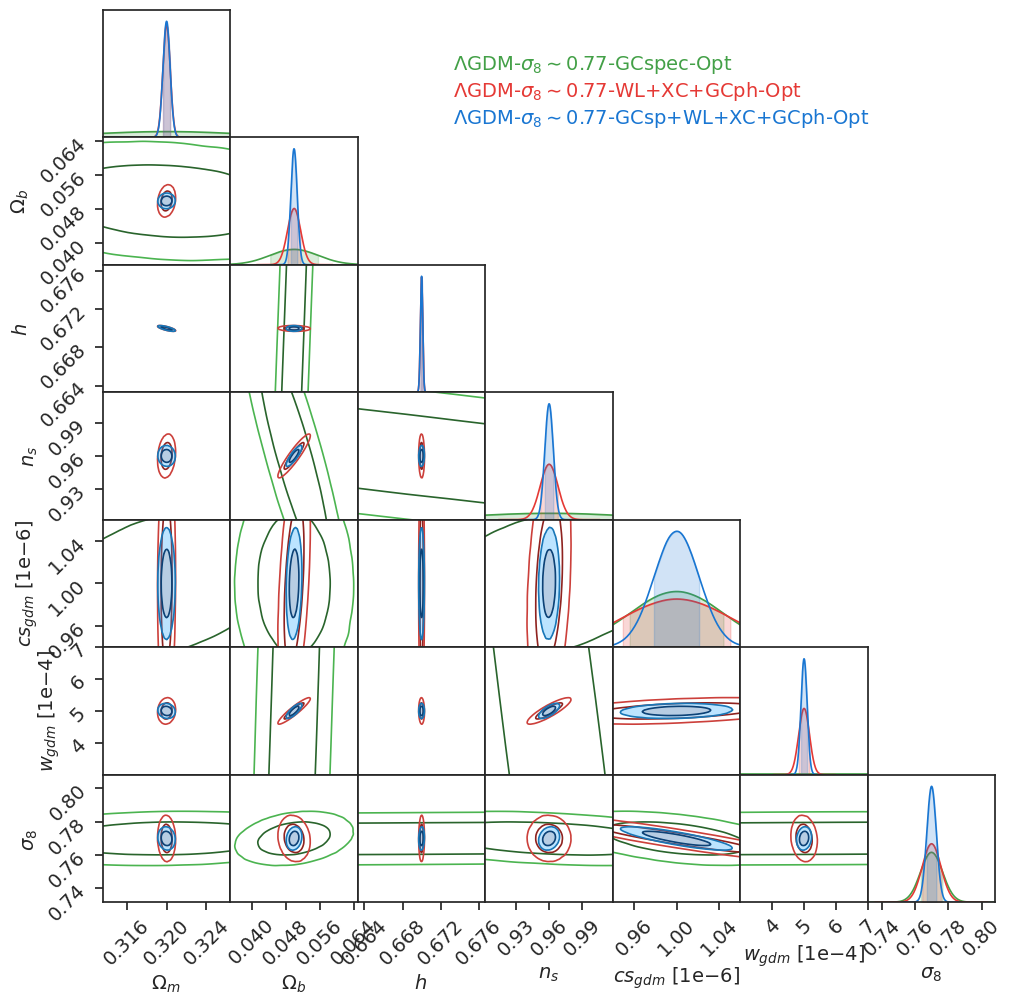}&
    \includegraphics[width=0.5\linewidth]{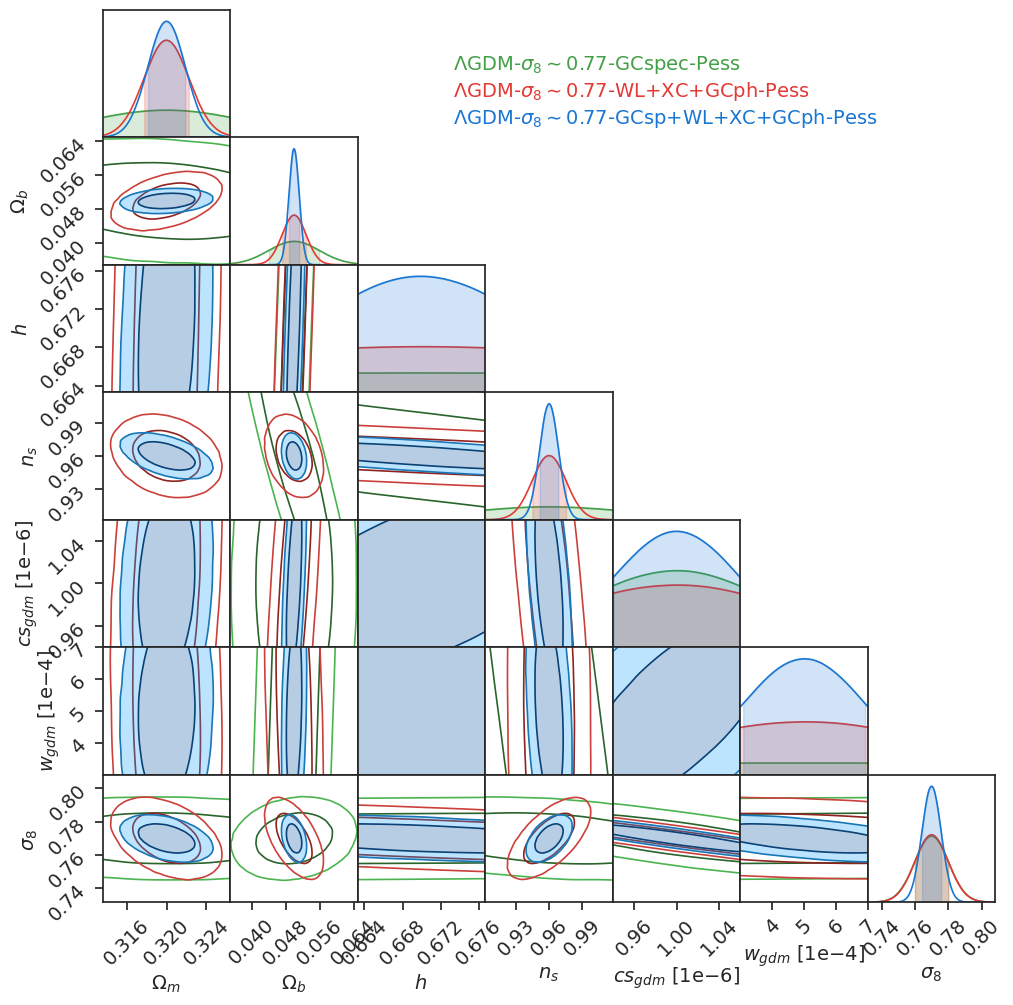}
    \end{tabular}
    \caption{\textbf{Left:} 1 and 2$\sigma$ joint marginal error contours on the cosmological parameters for GDM III model with from Fisher forecasts in the optimistic settings. In green using spectroscopic Galaxy Clustering (GCsp), in red the photometric probes and their cross-correlations (WL+GCph+XC) and in blue all the photometric probes with their cross correlation combined with spectroscopic Galaxy Clustering (GC$_{\rm sp}$+WL+GCph+XC).	
	  \textbf{Right:} 1 and 2$\sigma$ joint marginal error contours on the cosmological parameters for the same model with the same previous probe combinations but in the pessimistic settings.}
    \label{fig:077XCGC}
\end{figure*}

\begin{figure*}[htbp]
    \centering
    \includegraphics[width=0.55\columnwidth]{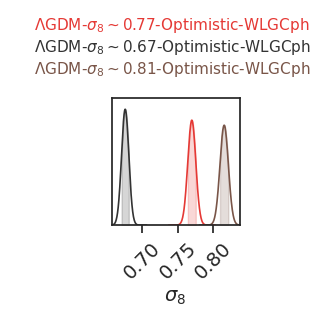}
    \includegraphics[width=0.55\columnwidth]{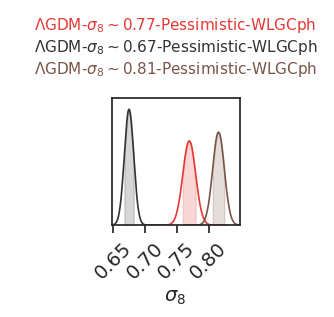}
    \includegraphics[width=0.55\columnwidth]{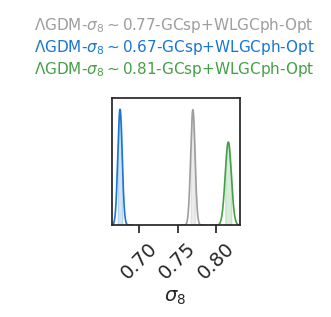}
    \includegraphics[width=0.55\columnwidth]{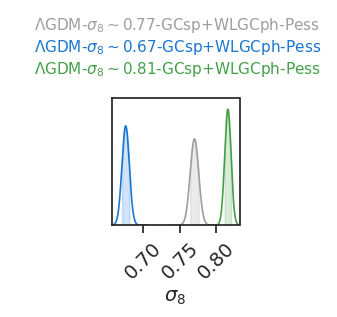}
    \centering
    \includegraphics[width=0.55\columnwidth]{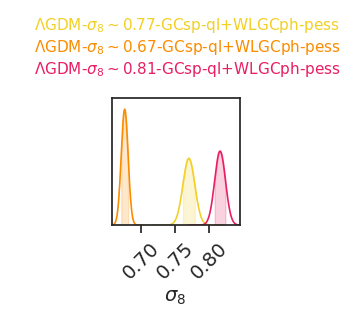}
    \caption{1 $\sigma$ joint marginal error contours on three adopted fiducial values for the $\sigma_8$ cosmological parameter for the GDM model with different combination of probes, or level of optimism considered (see the legend for details in each panel)}
    \label{fig:067vs077vs081}
\end{figure*}

In \Cref{fig:081XCGC}, we show the 1 and $2\sigma$ elliptical contours for the GCsp probe (green), the WL+GCph+XC combination (red), and the full Euclid-like set of probes GCsp+WL+GCph+XC (blue) for the cosmological and GDM extension parameters in the GDM I case. The analogous results for the GDM II and GDM III cases are presented in \Cref{fig:067XCGC} and \Cref{fig:077XCGC}, respectively. All numerical constraints, including those from GCsp under the quasi-linear scale assumptions, are reported in \Cref{tab:rel-errors-GDM}. Finally, in \Cref{fig:067vs077vs081} we compare the $1\sigma$ contours of the GDM I, GDM II, and GDM III scenarios across the optimistic, pessimistic, and quasi-linear scale configurations for the full GCsp+WL+GCph+XC combination.

\begin{table*}[htbp]
\caption{Forecast $1\sigma$ relative marginal errors on the cosmological and model parameters relative to their corresponding fiducial value for GDM I, GDM II and GDM III (see \Cref{sec:fisher} for details) in the pessimistic, quasi-linear and optimistic cases, using Euclid-like observations of GCsp, WL+XC+GCph\ and GCsp+WL+XC+GCph.
}
\begin{tabularx}{\textwidth}{Xccccccc}
\hline
\hline
 \multicolumn{8} {c} {{{\mbox{\textbf{\rm GDM I}}\,\,\, \boldsymbol{$\sigma_8=0.81$}}}} \\ 
 \hline
 & \multicolumn{1}{c}{$\Omega_{\rm m,0}$} & \multicolumn{1}{c}{$\Omega_{\rm b,0}$} & \multicolumn{1}{c}{$c^2_{s,{\rm gdm}}$} & \multicolumn{1}{c}{$w_{\rm gdm}$} & \multicolumn{1}{c}{$h$} & \multicolumn{1}{c}{$n_{\rm s}$} &  \multicolumn{1}{c}{$\sigma_{8}$} \\
\hline
\rowcolor{lavender(web)}\multicolumn{8} {l}{{{Pessimistic setting}}} \\ 
GCsp\ $(k_{\rm max} = 0.15\,h\,{\rm Mpc}^{-1})$ (quasi-linear)  &	5.93	\%	&	23.91	\%	&	98.77	\%	&	611.06	\%	&	20.89	\%	&	14.58	\%	&	14.58	\%	\\
GCsp\ $(k_{\rm max} = 0.25\,h\,{\rm Mpc}^{-1})$	&	2.39	\%	&	11.67	\%	&	8.33	\%	&	374.11	\%	&	10.08	\%	&	7.78	\%	&	1.24	\%	\\
WL+XC+GCph	&	0.67	\%	&	5.59	\%	&	9.67	\%	&	89.12	\%	&	4.01	\%	&	1.27	\%	&	1.10	\%	\\
GCsp+WL+XC+GCph	&	0.57	\%	&	2.29	\%	&	5.38	\%	&	38.03	\%	&	1.60	\%	&	0.83	\%	&	0.67	\%	\\
\hline
\rowcolor{lavender(web)}\multicolumn{8} {l}{{{Optimistic setting}}}  \\ 
GCsp\ $(k_{\rm max} = 0.3\,h\,{\rm Mpc}^{-1})$	&	2.16	\%	&	11.05	\%	&	4.07	\%	&	337.38	\%	&	9.37	\%	&	7.02	\%	&	0.82	\%	\\
WL+XC+GCph	&	0.11	\%	&	3.01	\%	&	5.01	\%	&	3.26	\%	&	0.02	\%	&	0.82	\%	&	0.72	\%	\\
GCsp+WL+XC+GCph	&	0.10	\%	&	1.46	\%	&	1.96	\%	&	1.85	\%	&	0.02	\%	&	0.38	\%	&	0.35	\%	\\
\hline
\hline
  \multicolumn{8} {c} {{{\mbox{\textbf{\rm GDM II}}\,\,\, \boldsymbol{$\sigma_8=0.67$}}}} \\ 
 \hline
 & \multicolumn{1}{c}{$\Omega_{\rm m,0}$} & \multicolumn{1}{c}{$\Omega_{\rm b,0}$} & \multicolumn{1}{c}{$c^2_{s,{\rm gdm}}$} & \multicolumn{1}{c}{$w_{\rm gdm}$} & \multicolumn{1}{c}{$h$} & \multicolumn{1}{c}{$n_{\rm s}$} &  \multicolumn{1}{c}{$\sigma_{8}$} \\
\hline
\rowcolor{lavender(web)}\multicolumn{8} {l}{{{Pessimistic setting}}} \\ 
GCsp\ $(k_{\rm max} = 0.15\,h\,{\rm Mpc}^{-1})$ (quasi-linear)  &	6.22	\%	&	24.14	\%	&	92.30	\%	&	637.66	\%	&	21.24	\%	&	15.02	\%	&	13.63	\%	\\
GCsp\ $(k_{\rm max} = 0.25\,h\,{\rm Mpc}^{-1})$	&	2.75	\%	&	12.28	\%	&	10.33	\%	&	397.47	\%	&	10.65	\%	&	8.20	\%	&	1.48	\%	\\
WL+XC+GCph	&	0.74	\%	&	5.66	\%	&	9.52	\%	&	91.48	\%	&	4.05	\%	&	1.30	\%	&	1.07	\%	\\
GCsp+WL+XC+GCph	&	0.62	\%	&	2.54	\%	&	6.02	\%	&	39.89	\%	&	1.73	\%	&	0.84	\%	&	0.74	\%	\\
\hline
\rowcolor{lavender(web)}\multicolumn{8} {l}{{{Optimistic setting}}}  \\ 
GCsp\ $(k_{\rm max} = 0.3\,h\,{\rm Mpc}^{-1})$	&	2.53	\%	&	11.63	\%	&	5.15	\%	&	364.73	\%	&	9.94	\%	&	7.51	\%	&	0.93	\%	\\
WL+XC+GCph	&	0.13	\%	&	3.12	\%	&	5.15	\%	&	3.43	\%	&	0.02	\%	&	0.84	\%	&	0.74	\%	\\
GCsp+WL+XC+GCph	&	0.13	\%	&	1.55	\%	&	2.45	\%	&	2.04	\%	&	0.02	\%	&	0.42	\%	&	0.42	\%	\\
\hline
\hline
  \multicolumn{8} {c} {{{\mbox{\textbf{\rm GDM III}}\,\,\, \boldsymbol{$\sigma_8=0.77$}}}} \\ 
 \hline
 & \multicolumn{1}{c}{$\Omega_{\rm m,0}$} & \multicolumn{1}{c}{$\Omega_{\rm b,0}$} & \multicolumn{1}{c}{$c^2_{s,{\rm gdm}}$} & \multicolumn{1}{c}{$w_{\rm gdm}$} & \multicolumn{1}{c}{$h$} & \multicolumn{1}{c}{$n_{\rm s}$} &  \multicolumn{1}{c}{$\sigma_{8}$} \\
\hline
\rowcolor{lavender(web)}\multicolumn{8} {l}{{{Pessimistic setting}}} \\ 
GCsp\ $(k_{\rm max} = 0.15\,h\,{\rm Mpc}^{-1})$ (quasi-linear)  &	6.02	\%	&	23.84	\%	&	96.27	\%	&	618.00	\%	&	20.89	\%	&	14.66	\%	&	14.22	\%	\\
GCsp\ $(k_{\rm max} = 0.25\,h\,{\rm Mpc}^{-1})$	&	2.39	\%	&	11.67	\%	&	8.33	\%	&	374.11	\%	&	10.08	\%	&	7.78	\%	&	1.32	\%	\\
WL+XC+GCph	&	0.67	\%	&	5.59	\%	&	9.67	\%	&	89.12	\%	&	4.01	\%	&	1.27	\%	&	1.16	\%	\\
GCsp+WL+XC+GCph	&	0.57	\%	&	2.29	\%	&	5.38	\%	&	38.03	\%	&	1.60	\%	&	0.83	\%	&	0.71	\%	\\
\hline
\rowcolor{lavender(web)}\multicolumn{8} {l}{{{Optimistic setting}}}  \\ 
GCsp\ $(k_{\rm max} = 0.3\,h\,{\rm Mpc}^{-1})$	&	2.16	\%	&	11.05	\%	&	4.07	\%	&	337.38	\%	&	9.37	\%	&	7.02	\%	&	0.87	\%	\\
WL+XC+GCph	&	0.11	\%	&	3.01	\%	&	5.01	\%	&	3.26	\%	&	0.02	\%	&	0.82	\%	&	0.76	\%	\\
GCsp+WL+XC+GCph	&	0.10	\%	&	1.46	\%	&	1.96	\%	&	1.85	\%	&	0.02	\%	&	0.38	\%	&	0.37	\%	\\
\hline
\hline
\end{tabularx}
\label{tab:rel-errors-GDM}
\end{table*}

We find that, across all levels of optimism and for all chosen models, the Euclid-like  spectroscopic survey alone is not sufficient to constrain the model extension parameter $w_{\rm gdm}$, particularly when restricting the analysis to linear scales. In contrast, the sound speed parameter $c^2_{s,{\rm gdm}}$ can be constrained to the $\sim5\%$ level, but only under the most optimistic survey assumptions.

This behavior can be understood from the fact that, in our GDM scenarios, both $w_{\rm gdm}$ and a fortiriori $c_{s,{\rm gdm}}^2$, leave only a small imprint on the background expansion. As a consequence, the AP signature in the GCsp probe is only weakly affected. The RSD contribution is more sensitive to GDM effects in the nonlinear regime, but it primarily constrains $c_{s,{\rm gdm}}^2$. The situation improves significantly when WL, GCph, and XC are combined, or when all \Euclid probes are used together (GCsp+WL+GCph+XC). In this case, both $c_{s,{\rm gdm}}^2$ and $w_{\rm gdm}$ can be constrained at the $\sim 5\%$--$1\%$ level under optimistic assumptions, with $w_{\rm gdm}$ even achieving tighter constraints than $c_{s,{\rm gdm}}^2$ in the full-probe combination. This is expected for $c_{s,{\rm gdm}}^2$ since it affects the growth factor of the matter power spectrum and benefits from the orthogonality of both ingredients. A similar improvement also arises for 
$w_{\rm gdm}$, given that is modifies the power spectrum at the nonlinear levels, as shown in \citet{Pace:2019vrs}.

The same trend is observed for the remaining cosmological parameters. In this case, there is a constraint at the $\sim 20\%$--$\sim 10\%$ level even with the spectroscopic GCsp alone and the pessimistic settings. This is expected since cosmological parameters do indeed affect the observables, as already shown in \citet{Blanchard:2019oqi}. The order-of-magnitude improvement obtained when adding the photometric probes or combining all probes arises from the breaking of degeneracies introduced when extending the model with the GDM parameters. This degeneracy break is visible in the near–orthogonality of the contours of GCsp and WL in \Cref{fig:081XCGC,fig:067XCGC,fig:077XCGC}, particularly for the combination of parameters $h$--$\Omega_m$, $h$--$\sigma_8$, and $h$--$w_{\rm gdm}$.
For the remaining ones, the improvement

For the remaining parameters, the improved or comparable performance of the WL+GCph+XC combination relative to GCsp alone for $\Omega_{\rm b}$, $h$, and $n_s$ arises from the strong constraining power of the 3$\times$2pt probes on $w_{\rm gdm}$ and, to a large extent, on $\Omega_{\rm m}$ and $\sigma_8$, which GCsp alone cannot constrain well, thus breaking the degeneracies introduced by the GDM extension.

All these results reflect the ability of \Euclid probes to distinguish between the different $\sigma_8$ normalizations adopted in this work. This is illustrated in \Cref{fig:067vs077vs081}, where we show the 1D confidence contours for the three $\sigma_8$ scenarios. Using the 3$\times$2pt probe alone (top row), the GDM-predicted value of $\sigma_8$ can already be distinguished from the CMB $\Lambda$CDM value in both optimistic and pessimistic settings, reaching a significance of about $3\sigma$ in the optimistic case, while some overlap remains in the pessimistic one.

Additionally, combining the 3$\times$2pt probes with GCsp (second row) allows us to detect the discrepancy at a clear $\sim 5\sigma$ level in the optimistic settings, and above $3\sigma$ in the pessimistic case. Only when adopting the most pessimistic GCsp configuration (bottom row) does the significance decrease again to around $3\sigma$. Hence we see that the combination of all our probes, even in the most conservative cases, will allow strong constraints and ruling out, either the GDM model, or confirm the discrepancy on $\sigma_8$.

\section{Conclusions} \label{sec:conclusions}

As done in \citet{Blanchard:2019oqi}, forecasts are computed for spectroscopic galaxy clustering (GCsp); photometric galaxy clustering (GCph), weak-lensing cosmic shear (WL) and the cross-correlation between the latter two probes (XC); or a combination of the previous two kind of LSS probes. The extension presented in this paper with respect to the forecasting pipeline of \citet{Blanchard:2019oqi} is twofold: first, due to the cosmologies beyond $\Lambda$CDM considered in this work; secondly, due to the modified equations entering the different probes:
\begin{enumerate}
    \item We have first computed the input quantities obtained from the linear and non linear evolution, finding broad agreement results for the redshifts and scales of interest between dedicated simulations and predictions from nonlinear recipes for the matter power spectrum.
    \item We then adapt the forecasting pipeline of \citet{Blanchard:2019oqi} to incorporate the additional modifications required by the $\Lambda$GDM scenario.    
\end{enumerate}

Hence, we have computed forecasts in optimistic and pessimistic scenarios, depending on the range of scales considered for each probe and on the level of systematics included. In the specific case of GCsp, we have also shown the impact of adopting a quasi-linear, more conservative choice, motivated by the additional uncertainties in the nonlinear predictions for a theoretical model beyond $\Lambda$CDM.

In an optimistic scenario, combining all \Euclid primary probes and considering our GDM~I fiducial model, whose value of $\sigma_8$ is close to that inferred from CMB experiments, we find that Euclid-like data would be able to constrain the GDM parameters $c^{2}_{s,{\rm gdm}}$ and $w_{\rm gdm}$ at the percent level using the full combination GCsp+WL+GCph+XC. The uncertainties increase slightly, to around $5\%$, when considering the photometric probes alone or the spectroscopic probe alone (the latter applying only to $c^{2}_{s,{\rm gdm}}$). Under more pessimistic settings, the constraints degrade to $\sim 5\%$ for the full combination and to $\sim 10\%$ when each probe class is used independently.

For the GDM~II fiducial model, whose $\sigma_8$ value is not favoured by current observations within the GDM framework, we obtain constraints of similar magnitude. This indicates that \Euclid{} observations should be able to distinguish between the GDM~I and GDM~II scenarios once data become available.

Finally, for the GDM~III fiducial model, whose $\sigma_8$ is closer to that preferred by low-redshift probes, we find constraints of comparable precision on $c^{2}_{s,{\rm gdm}}$ and $w_{\rm gdm}$, as well as on $\sigma_8$ itself. This implies that a \Euclid-like survey will be able to test whether the GDM framework can account for the observed discrepancy in the matter-fluctuation amplitude.

This highlights the significant amount of information that can be extracted from the cross-correlation of photometric probes, especially when a wider range of nonlinear scales is included. In conclusion, a Euclid-like survey, and by then \Euclid itself, will be able to provide outstanding constraints on extensions beyond the concordance model, with conservative and even more with optimistic cases showing its ability to probe nonlinear scales, where most of the cosmological information resides. However, achieving reliable final results will require accurate modeling of the theoretical observables at these scales, as done in this work, and careful control of systematic uncertainties once data become available.

\begin{acknowledgements}

The authors would like to acknowledge S. Srinivasan for helping with the validation tests. The authors also acknowledge fruitful discussions with D. B. Thomas, as well as useful critics from F. Pace and D. Bertacca. ZS acknowledges support from the research projects PID2021-123012NB-C43, PID2024-159420NB-C43, the Proyecto de Investigación SAFE25003 from the Consejo Superior de Investigaciones Científicas (CSIC), and the Spanish Research Agency (Agencia Estatal de Investigaci\'on) through the Grant IFT Centro de Excelencia Severo Ochoa No CEX2020-001007-S, funded by MCIN/AEI/10.13039/501100011033. JNLS acknowledges the support by the European Union and the Czech Ministry of Education, Youth and Sports (Project: MSCA Fellowships CZ FZU III -CZ$.02.01.01/00/22\_010/0008598$).

\end{acknowledgements}

\bibliographystyle{aa}
\bibliography{biblio}

\end{document}